\documentstyle[12pt]{article}

\newfont{\twelvemsb}{msbm10 scaled\magstep1} 
\newfont{\eightmsb}{msbm8} 
\newfam\msbfam 
\textfont\msbfam=\twelvemsb 
\scriptfont\msbfam=\eightmsb 
\catcode`\@=11 
\def\Bbb{\ifmmode\let\next\Bbb@\else 
  \def\next{\errmessage{Use \string\Bbb\space only in math mode}}\fi\next} 
\def\Bbb@#1{{\fam\msbfam{{#1}}}}

\newcommand{\be}{\begin{equation}} 
\newcommand{\ee}{\end{equation}} 
\newcommand{\ba}{\begin{eqnarray}} 
\newcommand{\ea}{\end{eqnarray}} 
\newcommand{\spz}{\hspace{0.2cm}}
\newcommand{\spZ}{\hspace{0.5cm}} 
\newcommand{\virg}{,\spz}
\newcommand{\virG}{,\spZ}

\def\d{\delta}

\def\t{\theta}

\newcommand{\la}{\lambda} 
\newcommand{\p}{\partial} 
\newcommand{\dx}{\partial_x} 
\newcommand{\dt}{\partial_t}

\newcommand{\nn}{\nonumber} 
\newcommand{\lt}{\left(} 
\newcommand{\rt}{\right)}

\newcommand{\bC}{{\Bbb C}}

\newcommand{\bZ}{{\Bbb Z}}

\newcommand{\cL}{{\cal L}}

\newcommand{\NP}[1]{Nucl.\ Phys.\ {\bf #1}} 
\newcommand{\PL}[1]{Phys.\ Lett.\ {\bf #1}} 
 
\newcommand{\CMP}[1]{Comm.\ Math.\ Phys.\ {\bf #1}} 
 
\newcommand{\PR}[1]{Phys.\ Rev.\ {\bf #1}} 
\newcommand{\PRL}[1]{Phys.\ Rev.\ Lett.\ {\bf #1}}

\begin{document} 
\sloppy 
\renewcommand{\thefootnote}{\fnsymbol{footnote}} 
 
\newpage 
\setcounter{page}{1}

\vspace{0.7cm} 
\begin{flushright} 
DCPT-01/71
\end{flushright} 
\vspace*{1cm} 
\begin{center} 
{\bf Aspects of Symmetry in Sine-Gordon Theory}
\footnote{Contribution to the Proceedings of Fourth International Conference
{\it Symmetry  in Nonlinear Mathematical Physics}, July 9-15, 2001, Kyiv, UKRAINE.
}\\ 
\vspace{1.8cm} 
{\large D.\ Fioravanti}\\ 
\vspace{.5cm} 
{\em Department of Mathematical Sciences, University of Durham,}\\
{\em South Road, DH1 3LE DURHAM, UNITED KINGDOM} \\
\vspace{.3cm}  
{\em e-mail: davide.fioravanti@dur.ac.uk}
\end{center} 

\vspace{1cm} 
 
\renewcommand{\thefootnote}{\arabic{footnote}} 
\setcounter{footnote}{0}

\begin{abstract}
{\noindent As} a prototype of powerful non-abelian symmetry in an Integrable
System, I will show the appearance of a Witt algebra of vector fields in the SG theory.
This symmetry does not share anything with the well-known Virasoro algebra
of the conformal $c=1$ unperturbed limit. Although it is quasi-local in the SG field
theory, nevertheless it gives rise to a local action on $N$-soliton solution variables.
I will explicitly write the action on special variables, which possess a
beautiful geometrical meaning and enter the Form Factor expressions of quantum theory.
At the end, I will also give some preliminary hints about the quantisation.
\end{abstract}

\vspace{1cm} 
 
 
\newpage

\section{Introduction}
Nowadays the very peculiar r\^ole of symmetries is clearly
recognised in all the areas of Mathematical Physics also thanks
to the recent developments of Quantum Physics. In fact, it was in
the context of Classical Physics that Liuoville defined as
integrable a system having a number of local integrals of motion
in involution (LIMI's) equal to the degrees of freedom and
proposed a theorem (Liouville-Arnold theorem \cite{Ar}) to {\it solve the
motion up to quadratures} -- in the case of finite number of
degrees of freedom. Nevertheless, there is no equivalent theorem
when the degrees of freedom become infinite as well as the number
of abelian symmetries: the classical field theories represent an
important example which attracted more and more interest. The
situation is even more complicated when the system is a quantum
field theory: in this case we may be interested, for instance, in
the energy spectrum \cite{Ba,KBI} or in the spectrum of fields or
in the correlation functions of those fields \cite{AlZFMS}, as
the usual meaning of motion is definitely lost. In fact, in
systems with infinite degrees of freedom non-abelian symmetries
revealed to be more useful: let us think of Classical Inverse
Scattering Method \cite{FT} and Bethe Ansatz \cite{Ba,KBI} as two
illustrative examples among the others. Moreover, the Virasoro
algebra in two Dimensional Conformal Field Theories (CFT's)
represents perhaps the most successful example of how a
non-abelian symmetry can {\it solve} a quantum field theory and
in this case a theory realising a physical system at the very
important {\it critical point} \cite{ISZ}.

Unfortunately, this Virasoro algebra do not exist any longer if
the system is pushed out of the critical point, still preserving
Liouville integrability \cite{ISZ}. For instance, the
Sine-Gordon (SG) theory is one of the simplest massive Integrable
Field Theories (IFT's), although it is the first theory in a
series of structure richer theories, the Affine Toda Field
Theories (ATFT's) \cite{DS} and possesses all the features
peculiar to the more general IFT's \cite{Col}. Actually, non-abelian
infinite-dimensional symmetries were found in all Toda theories
and they are called dressing symmetries at classical level
\cite{Sem} and become (level $0$) affine quantum algebras after
quantisation \cite{BLRS}. Nevetheless, because of their affine
and highly non-local characters those symmetries are not of large
use.

In this talk I present the appearance of infinitesimal symmetry
transformations (vector fields) acting on the boson field of the
classical Sine-Gordon theory. These vector fields turn out to
close a Witt (centerless Virasoro) algebra. Since the only {\it
ingredient of the recipe} is the Lax pair formulation of SG
equation, it is clear how to generalise the construction to more
general field theories like, for instance, ATFT's. Nevertheless,
I rather would like to focus my attention on the origin and form
of the infinitesimal transformations in the particular case of SG
theory. Specifically, I will show how to introduce the SG theory
starting from the simpler Korteweg-de Vries (KdV) theory and how
to frame this symmetry inside the KdV theory. Actually, I will not
give a complete proof of all the statements I will formulate,
leaving this part to a more systematic publication \cite{35}. On
the contrary, the restriction of these vector fields on the variables
of the $N$-soliton solutions was described and analysed in
\cite{FS2}: in this talk I sketch only how to derive this action
on a more intuitive ground. In the soliton phase space the
infinitesimal transformations are realised in a much simpler form
and in particular they become local contrary to the field theory
case (in which these are quasi-local). At the end, I will deliver 
few comments about how much
easier a quantisation of the soliton phase space might reveal.

\section{The action of the Witt symmetry on fields}
Let me recall the construction of the Witt symmetry in the context
of (m)KdV theory \cite{GO, FS}.  It was shown in \cite{FS}, following the
so-called matrix approach, that it appears as a generalisation of the
ordinary dressing transformations of integrable models. 
As integrable system the mKdV equation enjoys a zero-curvature 
representation
\be
[ \dt - A_t , \dx - A_x ] = 0 ,
\label{zcurv}
\ee
where the Lax connections $A_x$ , $A_t$ belong to a finite dimensional representation
of some loop algebra and contain the fields and their derivatives. In this particular 
case the first Lax operator ${\cal L}$ is given by
\be
A_x= \left(\begin{array}{cc} \phi' & \la \\
                              \la & -\phi' \end{array}\right),
\label{flaxc}
\ee
where I have denoted with $\phi'$ the mKdV field (prime means derivative with respect to 
the {\it space variable} $x$), with $\la$ the $A_1^{(1)}$ loop algebra parameter 
(spectral parameter) and $A_t$ can be found using the dressing procedure I am going 
to describe \cite{FS0}. The KdV variable $u(x)$ is connected to the mKdV field 
$\phi'$ by the Miura transformation:
\be
u=-(\phi')^2-\phi''.
\label{miura}
\ee
Key objects in the following construction are solutions
$T(x,\lambda)$ of the so-called associated linear problem
\be
(\partial_x-A_x(x,\lambda))T(x,\lambda)=0,
\label{aslin}
\ee
which may be called monodromy matrices.
A formal (suitably normalised) solution
of (\ref{aslin}) can be formally expressed by
\be
T_{reg}(x,\la) =  e^{H\phi(x)}{\cal P} \exp\lt\la \int_0^xdy
(e^{-2\phi(y)} E +e^{2\phi(y)} F )\rt .
\label{regsol}
\ee
Of course, this solution is just an infinite series
in positive powers of $\lambda\in \bC$ with an infinite radius of convergence. I
shall often refer to (\ref{regsol}) as {\it regular expansion}. It is also
clear from (\ref{regsol}) that any solution $T(x,\la)$ possesses an essential
singularity at $\la=\infty$ where it is governed by the
corresponding {\it asymptotic expansion}. In consequence, an asymptotic expansion 
has been derived in detail in \cite{FS0}
\be
T_{asy}(x,\la)=KG(x,\la)e^{-\int_0^x dy D(y)},
\label{asyexp}
\ee
where $K$ and $G$ and $D$ are written explicitly in \cite{FS0}. In
particular the matrix
\be
D(x,\la)=\sum_{i=-1}^{\infty}\la^{-i}d_i(x)H^i \virg
H= \left(\begin{array}{cc} 1 & 0 \\
                           0 & -1\end{array}\right)
\ee
contains the local
conserved densities $d_{2n+2}(x)$.

Obviously, a gauge transformation for $A_x$
\be
\delta A_x(x,\lambda)=[\t(x,\lambda),{\cal L}]
\label{gauge}
\ee
preserve the zero-curvature form (\ref{zcurv}) if an analogous one applies
to $A_t$: the vector field $\delta$ defines a symmetry of {\it the 
equation of motion} (\ref{zcurv})
in the usual sense, mapping a solution into another solution.
Moreover, to build up a consistent gauge connection $\t(x,\lambda)$ for the previous
infinitesimal transformation, I must pay attention to the fact
that the r.h.s. needs to be independent of $\la$ since the
l.h.s is, as consequence of (\ref{flaxc}). Hence, a suitable choice
for the gauge connection goes through the construction of the following object
\be
Z^X(x,\lambda)=T(x,\lambda)XT(x,\lambda)^{-1},
\label{zdress}
\ee
where $X$ is such that
\be
[\dx, X]=0.
\ee
Indeed, it is obvious from the previous definition that it satisfies the
{\it resolvent condition}
\be
[\cL,Z^X(x,\la)]=0,
\label{resolv}
\ee
for the first Lax operator $\cL=\partial_x-A_x(x,\lambda)$. Now, this
property implies
\be
[\cL,(Z^X(x,\la))_-]=-[\cL,(Z^X(x,\la))_+],
\ee
where the subscript $-$ ($+$) means that I restrict the series only to negative
(non-negative) powers of $\la$, and hence yields the construction of a
consistent gauge connection defined as
\be
\t^X(x,\la)=(Z^X(x,\la))_- \spZ  or \spZ  \t^X(x,\la)=(Z^X(x,\la))_+.
\label{gaugepar}
\ee
Further, I have to impose one more consistency condition implied by the
explicit form of $A_x$ (\ref{flaxc}), namely $\delta A_x$ must be diagonal
\be
\d^X A_x=H\d^X \phi' .
\label{selfcons}
\ee
This implies restrictions about the indices of the transformations \cite{FS1}.
After posing $T=T_{reg}$ I obtain the so-called dressing symmetries
\cite{FS0} and the indices are even for $X=H$ and odd for $X=E,F$. Instead,
after posing $T=T_{asy}$ I get for $X=H$ the commuting (m)KdV flows (or the (m)KdV
hierarchy), which define the different {\it time} $t_{2k+1}\virg k=0,1,2,\dots$ 
{\it evolutions} and in particular (\ref{zcurv}) with $t=t_3$ \cite{FS1}.

At this point I want to make an important observation. Let me consider the KdV
variable $x$ as a {\it space direction} $x_-$ of some more general system (and
$\p_-= \p_x$ as a space derivative). Let me introduce the {\it time} variable
$x_+$ through the corresponding evolution flow
\be
\p_+ = (\d^E_{-1}+\d^F_{-1}),
\ee
defined by a zero curvature condition of the form (\ref{gauge})
\be
\p_+ A_{x_-}(x_-,x_+;\lambda)=[\t_+(x_-,x_+;\lambda),{\cal L}].
\label{timeevol}
\ee
This specific $\t_+(x_-,x_+;\lambda)$ is derived using (\ref{gaugepar})
with the regular expansion. 
Then, it can be proved \cite{FS1} that the equation of motion
for $\phi$ becomes:
\be
\p_+\p_-\phi=2\sinh(2\phi),\spZ or \spz if \spz \phi\rightarrow i\phi ,\spZ
\p_+\p_-\phi=2sin(2\phi)
\label{sgeq}
\ee
i.e. the Sine-Gordon equation. As we will see later, this observation
will reveal very fruitful for my purpose since
it provides an introduction of Sine-Gordon dynamics
as a vector field in the powerful algebraic framework
of the KdV hierarchy and its symmetries. For instance, I 
obtain as simple by-product the fact that mKdV hierarchy is a
symmetry for SG equation. Of course, the
Hamiltonians -- given by the part of the dressing charges 
corresponding to the (\ref{timeevol}) and the higher flows
$\p_{2k+1}=\d^E_{-2k-1}+\d^F_{-2k-1}$ \cite{FS0} -- coincide with 
the well-known ones \cite{FT}.

Now, let me explain how the Witt symmetry appears in the KdV
system \cite{FS}.
The main idea is that one may dress not only the generators of the underlying
$A_1^{(1)}$ algebra but also an arbitrary differential operator in the spectral
parameter. I take for example $\lambda^{m+1}\partial_{\lambda}$ which are the 
well known vector fields of the diffeomorphisms of the unit circumference
and close a Witt algebra. Then I proceed in the same way as above defining the
resolvent associated to the circumference diffeomorphisms
\be
Z^V(x,\la)=T(x,\la)\p_\la T(x,\la)^{-1} .
\label{zvir}
\ee
When I consider the asymptotic case, i.e. I take $T=T_{asy}$ in
(\ref{zvir}), I obtain the non-negative Witt flows.
In general they are written in terms of recursive quasi-local
expressions $\alpha^V_{2m}(x) \virg m\geq 0$ as
\be
\d^V_{2m} \phi(x)= \alpha^V_{2m}(x),
\label{vir1}
\ee
where
\be
\alpha^V_0(x)=-x\phi'(x) \virG
\alpha^V_{2m+2}(x)=[-\phi'\dx^{-1}\phi'\dx + {1\over 4}\dx^2] \alpha^V_{2m}(x).
\label{alpha}
\ee
Let me highlight the appearance of the pseudodifferential operator
$\dx^{-1}$, acting on a function $f(x)$ as
\be
\dx^{-1} f(x)=\int^x dy f(y),
\ee
which is responsible (together with the form of the initial condition
(\ref{alpha})) for the non complete locality.
From these vector fields I can deduce the action on $u(x)$
using (\ref{miura})
\be
\d^V_{2m} u(x) =2 \dx \beta^V_{2m+1}(x)
\label{vir2}
\ee
again in terms of recursive quasi-local expressions $\beta^V_{2m-1}(x)$
\be
\beta^V_{-1}=-x \virG
\beta^V_{2m+1}(x)=[{1\over 2}(u+\dx^{-1}u\dx+{1\over 2}\dx^2)]\beta^V_{2m-1}(x) .
\label{beta}
\ee
For instance, the first two of (\ref{vir1}) can be written as
\ba
\d_0^V\phi=-x\phi' \virG
\delta_{2}^V  \phi= \frac{1}{2}\phi'(\dx^{-1}\phi'^2)-\frac{1}{2}\phi''-
\frac{1}{2}x[\frac{1}{2}\phi'''-(\phi')^3].
\label{vir1e}
\ea
The negative Witt transformations can also be built up by taking $T=T_{reg}$ in
(\ref{zvir}) in such a way to complete the algebra \cite{FS}. Unfortunately, those
vector fields do not act as gauge transformations on the SG equation
of motion (\ref{timeevol}), actually they are not true symmetry transformations
\cite{FS1}.

Nevertheless, thanks to the way (\ref{timeevol}) I have introduced
SG theory through, I am in the position to extend the (half) Witt
symmetry algebra jumping from (m)KdV to the SG theory.
For obvious reasons I will rename in the following the
KdV variable $u$ with
\be
u^-(x_-,x_+)=-(\p_-\phi(x_-,x_+))^2 - \p_-^2\phi(x_-,x_+).
\label{u-}
\ee
Hence, after looking at the symmetric r\^ole that
the derivatives $\p_-$ and $\p_+$ play in the Sine-Gordon equation, I can
obtain {\it the negative (m)KdV hierarchy}, acting on
the fields $\phi(x_-,x_+)$ and
\be
u^+(x_-,x_+)= -(\p_+\phi(x_-,x_+))^2 - \p_+^2\phi(x_-,x_+),
\label{u+}
\ee
in the same way as above but with the change of r\^oles 
$x_-\rightarrow x_+$
(and consequently $\p_-\rightarrow\p_+$).
Similarly, I obtain the other half of a Witt algebra by using the same
construction already showed, but with $x_-$ substituted by $x_+$.

Of course, it is not obvious at all that the two different
halves will recombine into a unique Witt algebra. 
Actually, even the first Witt vector field in the original
construction (\ref{vir1e}) needs a {\it symmetrising improvement} to leave
exactly invariant the zero curvature form of SG equation (\ref{timeevol}):
\be
\d_0^V\phi=-x_-\p_-\phi-x_+\p_+\phi .
\label{d0}
\ee
Nevertheless, I have
checked this statement brute force in the case
\be
[\delta^V_{2},\delta^V_{-2}]\phi=4\delta^V_0 \phi,
\label{vire}
\ee
and it works in a peculiar manner, simply using the transformation definitions
(\ref{d0}) and the second of (\ref{vir1e}). I would like to leave
for future publication the detailed explanation of how a complete proof of
this proposition may be elaborated along smart lines \cite{35}.

In conclusion, I have found an entire Witt algebra of transformations acting
as gauge symmetries on SG equation (\ref{timeevol}).
Moreover, I sketch now how the restriction of the action on soliton solution
phase space yields the result argued in \cite{FS2} following a slightly different
procedure.

\section{The Witt Symmetry acting on the soliton solution variables}
I start with a brief description of the well known soliton solutions of
SG equation and (m)KdV hierarchy in the infinite {\it times} formalism.
To see how a $N$-soliton solution can be parametrised, I need to go through
the expression of the so-called {\it tau-function}. This can be written
as a determinant
\be
\tau(X_1,...,X_N| B_1,...,B_N)=\det(1+V)
\label{tau}
\ee
where $V$ is a $N$x$N$ matrix
\be
V_{ij}=2{B_iX_i \over B_i+B_j} \virg i,j=1,...,N,
\label{vmatr}
\ee
and $X_i(\{t_{2k+1}\}|x_i,B_i)$ are exponential functions of
of all the {\it times} $\{t_{2k+1}\}\virg k \in \bZ$
(e.g. in the previous notation $t_{-1}=x_+, t_1=x_-, t_3=t$)
\be
X_i(\{t_{2k+1}\}|x_i,B_i)=x_i\exp(2\sum_{k=-\infty}^{+\infty}
B_i^{2k+1} t_{2k+1}).
\label{X}
\ee
The constant parameters $B_i$ and $x_i$ describe the soliton rapidities and positions respectively.
Now the SG or the mKdV field solution is expressed in a beautiful
unitary way as
\be
e^\phi={\tau_-\over \tau_+} ,
\label{phiintau}
\ee
where simply
\be
\tau_\pm=\tau(\{\pm X_i\}|\{B_i\}),
\label{taupm}
\ee
in the sense that after putting all the negative (positive) times to zero, I
end up with the $N$-soliton solution of the mKdV hierarchy (the negative mKdV hierarchy), 
whereas after the position to zero of all the times but $t_{-1}=x_+, t_1=x_-$, I end up with the 
$N$-soliton solution of the SG equation.

The main goal of this Section is to find the action of the Witt symmetry on
the $N$-soliton solution and this is more conveniently achieved introducing other 
variables $\{A_i,B_i\}$, expressed implicitly by the old variables
$\{X_i,B_i\}$ through the implicit formul\ae
\be
X_j\prod_{k\ne j}{B_j-B_k\over B_j+B_k}=\prod_{k=1}^N{B_j-A_k\over B_j+A_k}
\virg j=1,...,N .
\label{analytic}
\ee
In fact, the $\{A_i,B_i\}$ are the soliton limit of certain variables
describing the more general quasi-periodic finite-zone solutions of (m)KdV
\cite{Nov}, being the $B_i$ the limit of the branch points of the hyperelliptic
curve describing a particular solution and the $A_i$ the limit
of the zeroes of the so-called Baker-Akhiezer function
defined on the curve. Actually, even for the description of the quantum
physics of Form Factors these variables are apparently more natural
and suitable \cite{BBS}. Although, in terms of these variables
the tau functions have still a combersome form
\ba
\tau_+ = 2^N\prod_{j=1}^N B_j\{{\prod_{i<j}(A_i+A_j)\prod_{i<j}(B_i+B_j)\over
\prod_{i,j}(B_i+A_j)}\}, \nonumber \\
\tau_- = 2^N\prod_{j=1}^N A_j\{{\prod_{i<j}(A_i+A_j)\prod_{i<j}(B_i+B_j)\over
\prod_{i,j}(B_i+A_j)}\},
\ea
the SG (mKdV) field (\ref{phiintau}) enjoys a simple expression
\be
e^\phi=\prod_{j=1}^N{A_j\over B_j}.
\label{analyticphi}
\ee
In consequence, the two components of the stress-energy tensor (\ref{u-}) and (\ref{u+}) 
take a wieldy form as well
\be
u^-=-2(\sum_{j=1}^NA_j^2-\sum_{j=1}^NB_j^2) \virG
u^+=-2(\sum_{j=1}^NA_j^{-2}-\sum_{j=1}^NB_j^{-2}).
\label{analyticu}
\ee
Now I am in the position to restrict the Witt symmetry of
SG equation developed in the previous Section to the 
case of soliton solutions. Although these 
transformations have been derived in \cite{FS2}, here I will 
follow a more intuitive path, which
underlines the geometrical meaning of this symmetry. In other words 
our starting point consists in the 
transformations of the rapidities under the Witt symmetry: 
I do expect that they change 
the conformal structure of the Riemann surface describing the finite-zone solutions. 
Actually, in the (m)KdV theory the soliton limit 
of the Witt action on the Riemann surface reads simply \cite{GO}
\be
\d_{2n}B_i=B_i^{2n+1} \virg n\ge 0,
\label{btransf}
\ee
where I have forgotten the superscript $V$ for indicating the action on soliton variables.
Further, the action of negative transformations should not be different
\be
\d_{-2n}B_i=-B_i^{-2n+1} \virg n>0,
\label{negb}
\ee
save an additional $-$ sign in the r.h.s. \cite{FS2} which takes into account the Witt 
algebra commutation relations. 
I have to show now the transformations of
the $A_i$ variables as consequences of (\ref{btransf}) and (\ref{negb}) 
once applied to the
implicit map (\ref{analytic}) by using the expression (\ref{X}) of $X_i$ in
terms of $B_i$. The problem is simplified by the fact that I know from the 
field theory that the symmetry algebra is 
a Witt algebra, and hence I need to compute only
the transformations $\d_0$, $\d_{\pm 2}$ and $\d_{\pm 4}$, for the higher
vector fields are then furnished by commuting.
In this way it is evident why the Witt transformations become {\it local}
when restricted on the soliton solutions, though the transformations of 
$\phi$ and $u^{\pm}$ in the SG theory
are quasi-local. 
Actually, I think more natural and more
compact to express the Witt action on $A_i$ by using the {\it equations of
motion} of $A_i$ derived from (\ref{X}) and (\ref{analytic}), like for instance
\ba
\d_{-1}A_i = \p_+A_i=\prod_{j=1}^N{(A_i^2-B_j^2)\over B_j^2}\prod_{j\ne i}
{A_j^2\over (A_i^2-A_j^2)}, \hspace{0.3cm}
\d_1 A_i = \p_-A_i=\prod_{j=1}^N(A_i^2-B_j^2)\prod_{j\ne i}
{1\over(A_i^2-A_j^2)}, \hspace{0.3cm} \nn \\
\d_3A_i = 3(\sum_{j=1}^N B_j^2-\sum_{k\ne i}A_k^2)\p_-A_i, \hspace{0.3cm}
{1\over 5}\d_5A_i = (\sum_{j=1}^N B_j^4-\sum_{k\ne i}A_k^4)\p_-A_i -
\sum_{j\ne i}(A_i^2-A_j^2)\p_-A_i\p_-A_j \hspace{0.2cm} .
\label{kdvonsol}
\ea
In conclusion, the direct calculation is quite tiresome and 
I present here only few results:
\ba
\d_{-2}A_i = \frac{1}{3} x_+\d_{-3}A_i
-A_i^{-1}-\p_+A_i \sum_{j=1}^NA_j^{-1}- x_-\p_+A_i, \hspace{4.5cm}\nn \\
\d_{-4}A_i = \frac{1}{5} x_+\d_{-5}A_i-
A_i^{-3}-\{\sum_{j\neq i}^N{1\over A_i}({1\over A_i^2}-
{1\over A_j^2})+\sum_{j=1}^N{1\over A_j}\sum_{k=1}^N
{1\over B_k^2}\}\p_+A_i-x_-\d_{-3}A_i
\label{result1}
\ea
and for non-negative vector fields
\ba
\d_0 A_i = (x_-\p_- -x_+\p_+ +1)A_i \virG
\d_2 A_i ={1\over 3}x_-\d_3A_i+A_i^3-(\sum_{j=1}^N A_j)
\p_-A_i-x_+\p_-A_i, \nn \\
\d_4 A_i ={1\over 5}x_-\d_5A_i+A_i^5-\{\sum_{j\ne i}A_i(A_i^2-A_j^2)+
\sum_{j=1}^NA_j \sum_{k=1}^N B_k^2\}\p_-A_i-x_+\d_3A_i.\hspace{0.6cm}
\label{result2}
\ea

At this point, I need to carry out two important checks. First, I have to
calculate the commutators of the $\d_{2m}$ (with $m\in\Bbb Z$) with 
the light-come SG flow $\p_{\pm}$, acting on $A_i$. These are always zero 
and represent an equivalent way to express the symmetry action.
Second, I have to verify the
algebra of the $\d_{2m}$ (with $m\in\Bbb Z$) on $A_i$ and this is 
a very non trivial check for I have derived all the transformations 
(\ref{result1}) and
(\ref{result2}) from the Witt algebra on $B_i$, written in (\ref{btransf})
and (\ref{negb}), and from the implicit map (\ref{analytic}). Nevertheless 
the action on $A_i$ is again a representation of the Witt algebra:
\be
[\d_{2n},\d_{2m}]A_i=(2n-2m)\d_{2n+2m} A_i \virg n,m\in {\Bbb Z}.
\label{primevir}
\ee

\section{Comments about quantisation}
Of course, I might be interested in the quantum Sine-Gordon
theory. In the case of solitons there is a standard procedure: the
canonical quantisation of the $N$-soliton solutions. Indeed, let me 
introduce the variables canonically conjugated to the $A_i$:
\be
P_j=\prod_{k=1}^N{B_k-A_j\over B_k +A_j} \virg j=1...,N.
\ee
In these variables one can perform the canonical quantisation
of the $N$-soliton system introducing the deformed commutation relations between
the operators $\hat{A}_i$ and $\hat{P}_i$ :
\ba
\hat{P}_j \hat{A}_j &=& q \hat{A}_j  \hat{P}_j , \nn \\
\hat{P}_k \hat{A}_j &=& \hat{A}_j  \hat{P}_k \spz for \spz k\ne j ,
\ea
where $q=\exp (i\xi)$, $\xi={\pi\gamma\over \pi-\gamma}$ and $\gamma$ is the 
coupling constant of the SG theory. Understanding how the Witt symmetry is 
deformed after quantisation is a very seductive problem.

{\bf Acknowledgements}-- It is a pleasure for me to thank E. Corrigan
and particularly M. Stanishkov for interesting discussions and the Organisers of 
the Workshop for invitation and very cordial hospitality. Further, I thank  EPSRC for the 
fellowship GR/M66370. This work has been
partially realised through financial support of TMR Contract ERBFMRXCT960012.

\end{document}